**COVER SHEET**

*NOTE: This coversheet is intended for you to list your article title and author(s) name only
—this page will not appear on the Electronic Product.*

Title: **Effects of Sound Suppressors on Muzzle Velocity, Bullet Yaw, and Drag**

Authors: Elya Courtney[1], Roy Couvillion[2], Amy Courtney[3], and Michael Courtney[4]

    INQUIRIES TO:   Michael Courtney, Ph.D.
                           BTG Research
                           9574 Simon Lebleu Road
                           Lake Charles, LA 70607 USA

                           Tel: +1-864-359-7962
                           E-mail: Michael_Courtney@alum.mit.edu

---

[1] University of Georgia, Department of Chemistry, Athens, Georgia, USA
[2] Acadian Armament, Lafayette, Louisiana, USA
[3] Exponent Scientific and Engineering Consulting, Philadelphia, Pennsylvania, USA
[4] BTG Research, 9574 Simon Lebleu Road, Lake Charles, Louisiana 70607 USA

Little has been published regarding whether and how sound suppressors impact bullet flight, including velocity, bullet yaw, and drag. These parameters were compared for four different bullets fired from a .300 Winchester Magnum under four different muzzle conditions (no device and three different suppressors). While effects were not observed in all cases, results indicate that sound suppressors can have the effect of reducing bullet yaw and drag significantly, and can also have small effects on muzzle velocity. Results further suggest that bullets with a propensity to yaw demonstrate significant reductions in yaw and drag when shot through a two stage symmetric suppressor versus unsuppressed or with a conventional mouse-hole/K-baffle design.

**INTRODUCTION**

Most studies on sound suppressors for small arms have focused on reductions in sound intensity [1, 2, 3]. Since the transition to baffle-based designs, which do not touch the projectile [3], it has often been assumed that effects of sound suppressors on bullet flight are insignificant. In the present study, effects of sound suppressors on bullet flight were investigated; results show that suppressor effects on muzzle velocity, drag, and yaw may not always be insignificant.

There are anecdotal reports of suppressor effects on accuracy and flight dynamics, as well as untested hypotheses regarding the responsible mechanisms. Transitional ballistics is not well understood in general, but it is believed in most cases (including suppressors), that unequal pressures at different points on the projectile can affect flight dynamics by introducing inaccuracy, yaw, and velocity variations. There is some concern that suppressors may also affect these aspects of bullet flight. It has been shown in larger guns that uneven pressure distributions near the muzzle are well correlated with the peak projectile yaw early in flight [4]. A common question is whether and to what degree suppressors influence bullet yaw, which would also influence bullet drag, since the total drag coefficient has a quadratic dependence on the angle of attack [5, Eq. 1].

Sound suppressors for small arms employ different engineering approaches for sound reduction. A two-stage suppressor manages the firearm muzzle blast in two distinct steps: 1) it contains the high-pressure discharge following the bullet, and 2) it slows the release of the gases to the atmosphere. The two tasks are managed sequentially within each of two separate volumes connected by a single aperture. In contrast, a conventional suppressor manages these tasks simultaneously and progressively through a series of chambers defined by baffles segmenting the volume of the suppressor. The "K-baffle" is a popular design so named by its cross-sectional geometry. The "mouse-hole" in the baffle adjacent to the bullet path hole is intended to enhance performance; however, it breaks the cylindrical symmetry of the suppressor, thus increasing potential for an uneven pressure distribution. The effects, if any, of these different designs on transitional and external ballistics have not been reported.

**METHOD**

Four muzzle attachment conditions were selected for testing: 1) no device 2) the Predator Cougar 8" two-stage design (Acadian Armament, Lafayette, LA) 3) the Predator Cougar 6" two-stage design (Acadian Armament, Lafayette, LA) and 4) the Silencerco Omega 30 K-baffle design (West Valley City, UT ). As is common in yaw card studies, "yaw" is used here to mean angle of attack, which technically includes contributions from both pitch and yaw angles in 6 degree of freedom (DOF) models. The rifle used in this study was a factory Remington 700 with 24" barrel chambered in 300 Winchester Magnum with a 1 in 10" twist. The barrel was carefully threaded for muzzle attachments indexing on the bore rather than the outer diameter of the barrel to maintain concentricity between the bore and the muzzle attachment to 0.002" or better.

Before completing the experimental design incorporating Doppler radar for velocity and drag measurements, a pilot study (15-30 shots with each muzzle condition) was performed using yaw cards from 45.7 m to 91.44 m with the four muzzle conditions, a 7.62x51 mm NATO precision rifle, and the 168 grain Sierra MatchKing (SMK) bullet loaded by Black Hills Ammunition (West Valley City, UT). A specially designed fixture was used to ensure that cards were normal to the flight direction to within 0.1°. After shooting, yaw cards were digitized with a resolution of 600 pixels per inch. Digital images were analyzed with ImageJ (Version 1.49f, National Institutes of Health) by fitting the perimeter of each bullet hole with an ellipse to determine the major and minor axes, which were then used with the length of the bullet bearing surface to estimate the bullet yaw when the bullet penetrated the card.

Results of the pilot yaw card study are summarized in Figure 1. Results were suggestive that suppressors have some small effects on yaw, but the largest yaw measured was under 3°, most yaws measured were under 2°, the average yaws were all between 1° and 2°, and the uncertainties did not provide a high level of statistical confidence. Such small yaw angles were unlikely to have significant effects on accuracy or drag, and would likely be near the limit of the Doppler radar ability to quantify drag differences (about 1%).

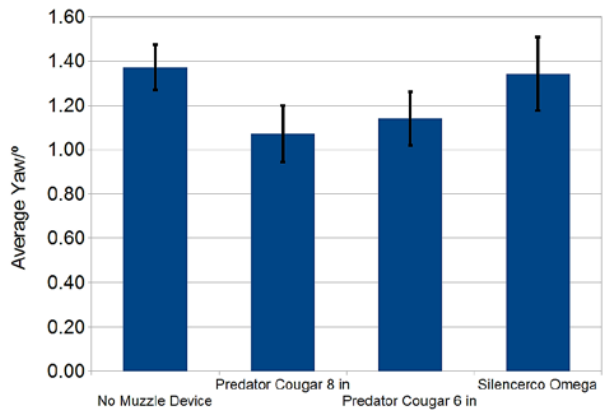

Figure 1. Average yaw from a 7.62x51 mm NATO rifle shooting the 168 SMK through four muzzle conditions.

Consequently, for the main experiment, the 300 Winchester Magnum and longer, heavier bullets were used to optimize the ability to detect any yaw effects, based on reasoning that 1) higher muzzle pressures would tend to create greater force imbalances in the transition event, 2) force imbalances on longer bullets would tend to create larger torques inducing larger yaw, and 3) longer bullets would have larger drag increases for the same yaw angles due to larger areas perpendicular to the velocity vector.

The Doppler radar measurement system for measuring free flight drag coefficients with an accuracy close to 1% has been described previously [6]. Briefly, a LabRadar unit (www.mylabradar.com ) is used to measure Doppler velocities at regularly spaced intervals from the muzzle out to 91.44 m (100 yards). Previous work has shown that bullet pitch and yaw damp out quickly over the first 91.44 m and that increases in drag due to yaw are relatively easily measured over the first 45.72 m (50 yards) [7, 8]. Air density was computed with the JBM ballistic calculator (www.jbmballistics.com) using ambient temperature, pressure, and relative humidity measurements measured with a Kestrel 4500 weather meter. Drag coefficients were then computed using Eq. 4 of Courtney et al. [9].

Reported muzzle velocities and drag coefficients were determined as the mean values from 5-10 shots each using four different factory loaded bullets fired under each of the four different muzzle conditions described above. The factory loaded bullets were the 200 grain Hornady ELD-X (Hornady Inc., Grand Island, NE), 200 ELD-X; the 190 grain AccuBond Long Range (Nosler, Inc., Bend, OR), 190 ABLR; the 195 grain Hornady Boattail Hollow Point, 195 H BTHP; and the 190 grain Sierra MatchKing (Sierra, Inc., Sedalia, MO), 190 SMK, loaded in .300 Winchester Magnum by Black Hills Ammunition (West Valley City, UT).

**RESULTS**

Mean muzzle velocities and their uncertainties are shown in Figure 2 for all test conditions. Muzzle velocities were as expected for commercial loads near full pressure for the SAAMI specifications in a 24" barrel for the given bullet weights.

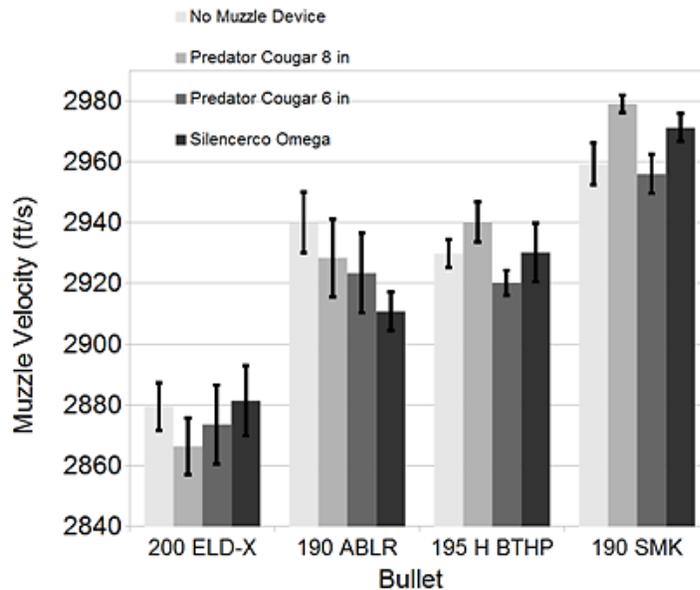

Fig. 2. Mean muzzle velocities for the four test bullets shot from .300 Winchester Magnum with four different muzzle conditions. Error bars represent the Standard Error of the Mean (SEM).

Muzzle velocities for the 200 ELD-X were all within the error bars of each other for all four muzzle conditions, suggesting there were no significant velocity variations with different muzzle devices for that load. In contrast, there was a decrease of about 30 ft/s (about 1%) between no muzzle device and the Silencerco Omega for the 190 ABLR load. There was also a small (but possibly significant) increase in muzzle velocities for the Predator Cougar 8" suppressor for the 195 H BTHP and the 190 SMK compared to other muzzle conditions. Rather than speculate on causes, we simply observe that these changes in muzzle velocities may be large enough to warrant care when predicting long-range trajectories. Specifically, the muzzle velocity used in predictions should be measured with the same muzzle condition for which an accurate trajectory calculation is needed.

Mean drag coefficients measured for each of the bullets under each of the test conditions are shown in Figure 3. Drag coefficients for the 200 ELD-X and the 190 SMK are consistent with the ballistic coefficients reported by the bullet manufacturers. Drag coefficients for the 190 ABLR are 5-10% higher than would be expected from the ballistic coefficient reported by Nosler, but this is not surprising since independent parties regularly report 5-10% more measured drag than claimed by Nosler [10].

In many cases for these three bullets, the error bars for different muzzle conditions tend to overlap each other, suggesting the differences in drag coefficients are not significant. However, the drag coefficient for the 190 ABLR through the Predator Cougar 8" model with symmetric two stage sound suppression is significantly smaller (4% or so) than the other three muzzle conditions. Further, the drag coefficient for the 190 SMK through the Predator 6" model of similar design is 2% smaller than with no device. These small drag differences may be significant due to the relative uncertainties that were often below 1%.

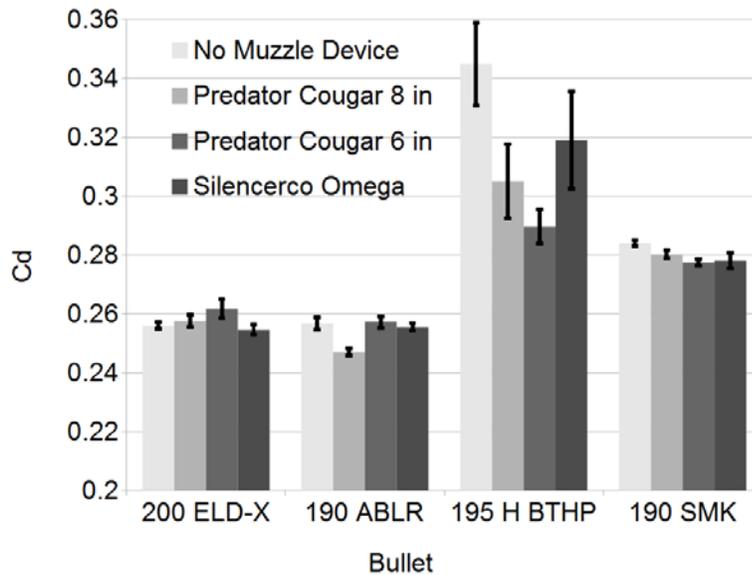

Figure 3. Mean drag coefficients for four test bullets shot from .300 Winchester Magnum with four different muzzle conditions. Error bars represent the Standard Error of the Mean (SEM).

    Both drag coefficients and error bars are much larger for the 195 H BTHP. The smaller drag differences for other bullets are difficult to attribute to yaw with confidence, but for the 195 H BTHP, both the increases in drag and the larger shot-to-shot drag variations seem attributable to yaw for several reasons. This bullet demonstrated noticeable decreases in drag from 45.72 m to 91.44 m compared with the first 45.72 m from the muzzle. This would be expected as the large yaw at shorter ranges damps out over increasing distance. Second, the raw Doppler data (V vs. t, not shown) reveals the expected oscillatory behavior in its slope expected from bullet coning motions. Third, shots with increased drag were noted to hit significantly further from the point of aim than shots with smaller drag. For whatever reason, this bullet tended to show significantly more yaw than the other three bullets in the study, and this large yaw resulted in large increases in bullet drag.

    Some effort was made to see how clearly the fast (nutation) and slow (precession) frequencies of the 195 H BTHP coning motion could be determined from the raw V vs. t data provided by the Doppler Radar. In 80% of the Fourier transforms, a possible slow coning (precession) frequency could be identified between 68 Hz and 77 Hz. Similarly, a possible fast coning (nutation) frequency could be identified between 243 Hz and 250 Hz for 80% of shots. The signal-to-noise ratio of the oscillations is not sufficient to quantify these frequencies more accurately with the available data, but a larger number of shots would likely provide a clearer view.

    The increased drag of the 195 H BTHP is not the same for all four muzzle conditions. Drag differences are not significant in all cases, but the drag is rank ordered largest to smallest for no device, the Omega suppressor (K-baffle with mouse hole), the Cougar 8" (two stage symmetric), and the Cougar 6" (two stage symmetric). The smaller uncertainty for the Cougar 6" suppressor results from smaller shot-to-shot variations and the difference between the mean drag coefficient for this suppressor and no muzzle device is statistically significant.

# DISCUSSION

This may be the first published report comparing transition and external ballistics of these two types of sound suppressors. The relationship between yaw and drag is long established [5, 11], but possible effects of modern suppressor design on bullet velocity and yaw have not been widely reported. Effects on muzzle velocity are small. Effects on yaw and drag also seem to be small in cases where the bullet is not prone to yaw. However, for bullets prone to yaw, the results presented here suggest that a two-stage muzzle device can reduce drag presumably resulting from transition-induced yaw.

Since the drag increase can be quantified with reasonable accuracy, the peak yaw angle could be estimated if the quadratic yaw drag coefficient were known. However, bullet manufacturers tend to keep this information proprietary, and quadratic drag coefficients have only been released for a few match style 7.62 mm bullets manufactured by Sierra [5]. Even though we observe a 2.3% reduction in drag for the 190 SMK with the 6 inch, two-stage symmetric suppressor, the yaw drag coefficient is unknown above M2.2 for this bullet. Extrapolating from Figure 22 in McCoy [5] suggests a negative yaw drag coefficient for experimental velocities here, and that is unwarranted.

One might consider whether detectable drag related yaw effects may be attributed to insufficient bullet stability for the 190 ABLR, 190 SMK and the 195 H HPBT. However, all bullets tested have gyroscopic stability over 1.5 at the muzzle velocities recorded from a 1 in 10" twist barrel and the ambient atmospheric conditions. Yaw is believed to arise from some transition effect causing a significant initial tip off rate rather than inadequate stability. A faster twist would change the precession and nutation dynamics, but it would not likely eliminate the larger peak yaw through increased stability.

In summary, test results support that yaw-related increases in drag are present in certain bullets that are prone to yaw. The methods used are able to support or refute anecdotal reports of the effects of muzzle devices by quantifying bullet yaw and drag over transitional and near ranges where the effects are greatest. Additional studies may inform whether certain classes of bullets are more prone to this effect and where the ballistic performance may benefit from use of a muzzle device. Future work might also consider additional muzzle device designs to better identify specific design features contributing to amelioration of transition-induced yaw.

**Acknowledgments**

The authors are grateful to the Fusilier Complex (Louisiana) for use of their test facilities. This work was funded by Acadian Armament and BTG Research; affiliated authors participated in all stages of the study with scientific and academic freedom.